\documentstyle[twocolumn,psfig,aps]{revtex} 

\newcommand{\rem}[1]{}
\newcommand{\authtwo}[4]{#1,#2,and~#3,#4}
\newcommand{\auththr}[6]{#1,#2,#3,#4,and~#5,#6}

\newcommand{\private}[2]{ (private communication).}

\newcommand{\yjfm}[5]{  { J. Fluid Mech. }{\bf #2}, #3 (#1).}

\newcommand{\yprl}[5]{  { Phys. Rev. Lett. }{\bf #2}, #3 (#1).}
\newcommand{\ypre}[5]{  { Phys. Rev. E. }{\bf #2}, #3 (#1).}

\newcommand{\ypf}[5]{  { Phys. Fluids }{\bf #2}, #3 (#1).}
\newcommand{\ypfa}[5]{  { Phys. Fluids A}{\bf #2}, #3 (#1).}

\newcommand{\yanf}[5]{  { Ann. Rev. Fluid Dyn. }
{\bf #2}, #3 (#1).}

\newcommand{\yphyd}[5]{  { Physica D} {\bf #2}, #3 (#1).}
\newcommand{\yjour}[6]{ { #2} {\bf #3}, #4 (#1).}

\newcommand{\sepl}[2]{  { Europhys. Lett. } (submitted).}
\newcommand{\sana}[2]{  { Astron. Astrophys. } (submitted).}
\newcommand{\tana}[2]{  { Astron. Astrophys. } (to be submitted).}
\newcommand{\pana}[2]{  { Astron. Astrophys. } (to be
published).}
\newcommand{\pmn}[2]{  { Monthly Notices Roy. Astron. Soc. }
(in press).}
\newcommand{\sjfm}[2]{  { J. Fluid Mech. } (submitted).}
\newcommand{\pjfm}[2]{  { J. Fluid Mech. } (to be published).}

\newcommand{\sprl}[2]{  { Phys. Rev. Lett. } (submitted).}
\newcommand{\pprl}[2]{  { Phys. Rev. Lett. } (to be
published).}
\newcommand{\spre}[2]{  { Phys. Rev. E } (submitted).}
\newcommand{\ppre}[2]{  { Phys. Rev. E } (to be published).}
\newcommand{\sprsl}[2]{  { Proc. Roy. Soc. Lond. } (submitted).}
\newcommand{\pprsl}[2]{  { Proc. Roy. Soc. Lond. } (in press).}
\newcommand{\spf}[2]{  { Phys. Fluids } (submitted).}
\newcommand{\ppf}[2]{  { Phys. Fluids } (to be published).}
\newcommand{\spp}[2]{  { Phys. Plasmas } (submitted).}
\newcommand{\ppp}[2]{  { Phys. Plasmas } (in press).}
\newcommand{\sgafd}[2]{  { Geophys. Astrophys. Fluid Dyn. } (submitted).}
\newcommand{\smn}[2]{  { Monthly Notices Roy. Astron. Soc. }
(submitted).}
\newcommand{\sapj}[2]{  { Astrophys. J. } (submitted).}
\newcommand{\papj}[2]{  { Astrophys. J. } (to be published).}
\newcommand{\ssph}[2]{  { Solar Phys. } (submitted).}
\newcommand{\psph}[2]{  { Solar Phys. } (to be published).}

\begin{document}
\title{ Transient vortex events in the initial value problem for turbulence}
\author{
D. D. Holm, T-Division and CNLS, MS-B284,
Los Alamos National Laboratory, Los Alamos, NM 87545, USA
\\{\footnotesize email: dholm@lanl.gov}
\and\\
Robert Kerr, Department of Mathematics,
University of Arizona, Tucson, AZ 85721, USA
\\{\footnotesize email: kerr@math.arizona.edu}
}
\date{October 18, 2001}
\maketitle

\begin{abstract}
A vorticity surge event that could be a paradigm for a wide class of
bursting events in turbulence is studied
to examine the role it plays in how the energy cascade is established.
The identification of a new coherent mechanism is suggested by
the discovery of locally transverse vortex configurations that are
intrinsically helical.  These appear simultaneously with
strong, transient oscillations in the helicity wavenumber co-spectrum.
At no time are non-helical, anti-parallel vorticity elements observed.
The new mechanism complements the traditional expectation that 
the development of a peak of the maximum vorticity $\|\omega\|_\infty(t)$ 
would be connected to nearly simultaneous growth of the dissipation,
eventually leading to the formation of the energy cascade with signatures 
such as spectra approaching -5/3 and strongly Beltramized vortex tubes.
Comparing how different large-eddy simulations treat the spectral transport of
helicity demonstrates that the dynamics leading
to the helical vortex configurations requires both
nonlinear transport and dissipation.  This finding emphasizes 
the importance of properly modeling both nonlinear 
transport and dissipation in large-eddy simulations.

\end{abstract}
\pacs{PACS number: 47.27.Cn,47.27.Eq}

\vspace{-20mm}
\begin{figure}[htbp]
\centering 
\hspace{-3cm}\psfig{file=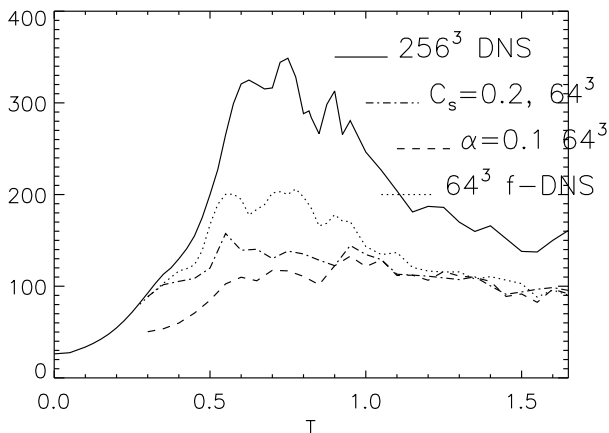,height=6.5cm}
\caption[]{
Comparison of the growth of $\|\omega\|_\infty$ between
our $256^3$ direct numerical simulation and three
results on $64^3$ meshes.  The full $256^3$ DNS filtered onto a $64^3$ mesh,
labeled $64^3$ f-DNS, a $64^3$ traditional Smagorinsky eddy viscosity
model, and a calculation using the LANS-$\alpha$ model.
This figure sets the time scales for our analysis.  
}
\label{vorticity-time}
\end{figure}

Although three-dimensional turbulence is characterized by
intermittent events both in space and time, it is
often envisaged as a homogeneous, statistically-steady
tangle of vortex tubes accompanied by a steady transfer of energy
through the spectrum to high wavenumbers and a $k^{-5/3}$ energy
spectrum. This statistically-steady description has been used to
verify turbulence models through simulations of forced turbulence and
decaying homogeneous, isotropic turbulence
\cite{clark79,\rem{mcmill79,lund91,Chasnov91,}Chen-etal[1999c]}.  
In these studies,
it is sometimes argued that the initial nonequilibrium transients can
be ignored as being non-universal.
However, each intermittent burst of turbulence is itself
a transient dynamical process involving individual vortex interactions
whose ensemble average is responsible for the overall statistical properties.
Our objective is to apply
a coordinated set of diagnostics in both physical and
wavenumber space for numerically detecting individual
vortex surge events and their effects upon the subsequent dynamics.

The numerical investigation of the decay of turbulence
discussed here is designed to improve our understanding
of individual intermittent events and continues a long tradition in 
numerical modeling of the initial value problems in
computational fluid dynamics. 
Numerically, the question of transient phenomena in turbulence
has been considered previously using a variety of initial conditions
\cite{Brachetetal[1983],KidaMurakami[1987],Herring-Kerr[1993],Kerr[1993],KidaTakaoka[1994]}.
In this letter the evolution from 
smooth, random, initial conditions introduced in
\cite{Herring-Kerr[1993]} to steady turbulent decay is considered 
using this coordinated set of diagnostics.
These diagnostics show that this initial value problem for turbulence
evolves through several complex states in a 
sequence of transitions.  These include:
\begin{description}
\vspace{-2mm}\item$\quad\bullet$
formation of vortex sheets that interact, encounter each other transversely
and then begin to roll up into vortex tubes,
\vspace{-2mm}\item$\quad\bullet$
development of a peak in the maximum vorticity $\|\omega\|_\infty(t)$
that is correlated with
helicity signatures in both physical and wavenumber space,
\vspace{-2mm}\item$\quad\bullet$
rearrangement of vortex tubes into transverse pairs 
having oppositely signed helicity 
$\Lambda=\mathbf{u}\cdot\omega$, where
vorticity $\omega={\rm curl}\mathbf{u}$.  Unlike
inviscid configurations\cite{KidaTakaoka[1994]}, the velocity $\mathbf{u}$ 
on each tube of the pair is only partially induced by its partner.  
Instead, $\mathbf{u}$ on the tubes arises primarily as a response to
strains and dissipation distributed within the pair.
\vspace{-2mm}\item$\quad\bullet$
formation of the classical decay regime with a $k^{-5/3}$ energy spectrum.
\end{description}
\vspace{-2mm}

\begin{figure}[htbp]
\psfig{file=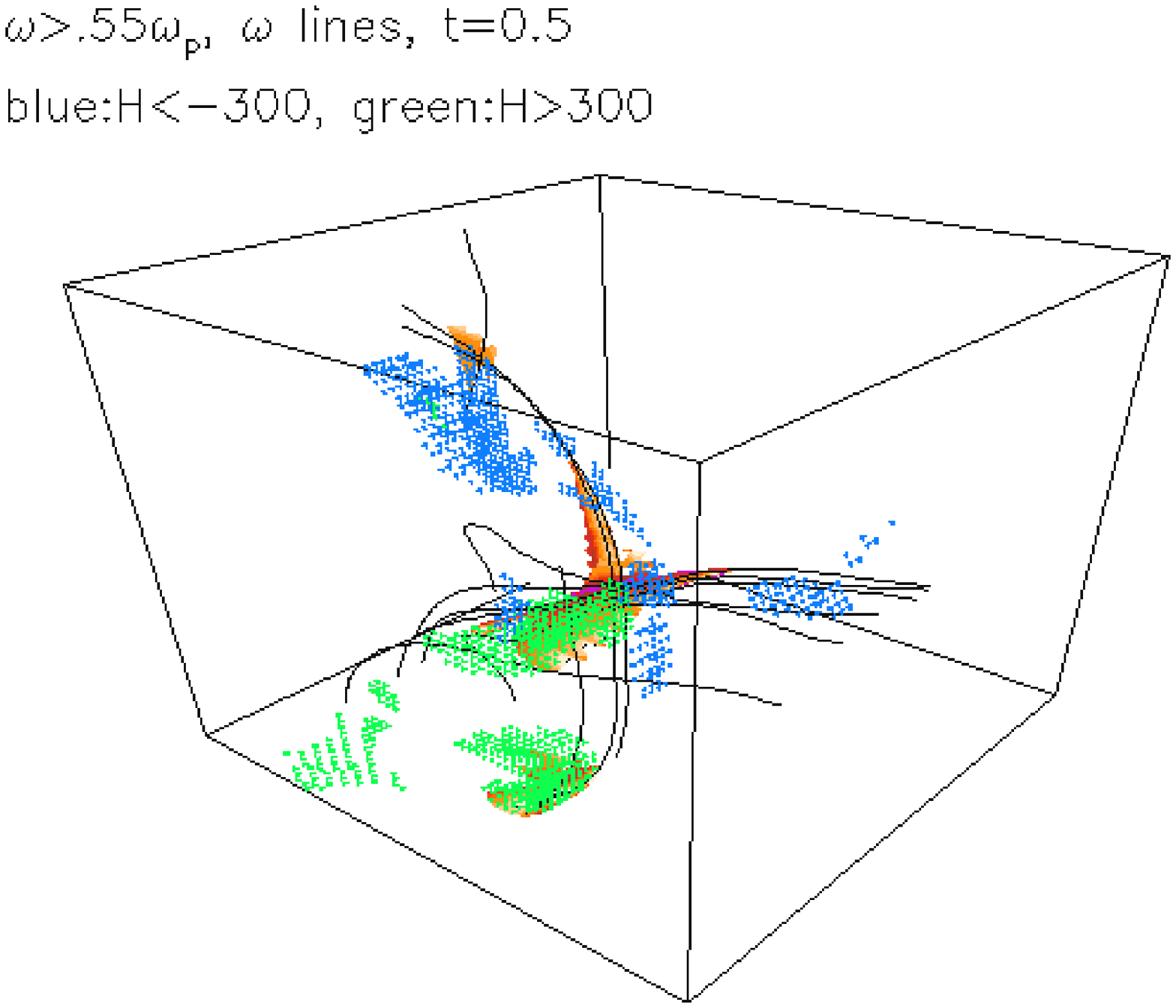,height=6.5cm}
\caption[]{Isosurfaces of vorticity at $t=0.5$ where
$\omega\geq0.55\|\omega\|_\infty$ in red/earth colors
with sample vortex lines through these regions.
Regions of high positive and negative helicity are indicated
by green and blue respectively.  The vortex lines meet transversely,
which is an inherently helical configuration.
}
\label{orthogonal-tubes}
\end{figure}

\begin{figure}[htbp]
\psfig{file=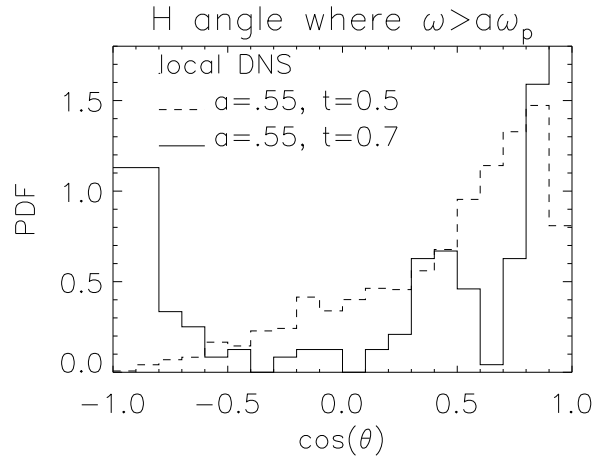,height=6.5cm}
\caption[]{Probability density functions of the helicity
angle $\cos\theta=\overrightarrow{u}\cdot\overrightarrow{\omega}/
(u\,\omega)$ in physical space
at $t=0.5$, during the surge in the vorticity
maximum, and $t=0.7$ after formation of the first distinct vortex tubes.
The distribution is taken over those points with vorticity
above the shown threshold, 55\% of the maximum vorticity 
$\omega_p=\|\omega\|_\infty$, in the
subdomain containing $\|\omega\|_\infty$ at both t=0.5 and 0.7.
The distribution was initially flat.  The asymmetry in the
distribution emerges because the transverse vortex configurations
that develop are inherently helical.
}
\label{helicity-pdf}
\end{figure}

\begin{figure}[htbp]
\psfig{file=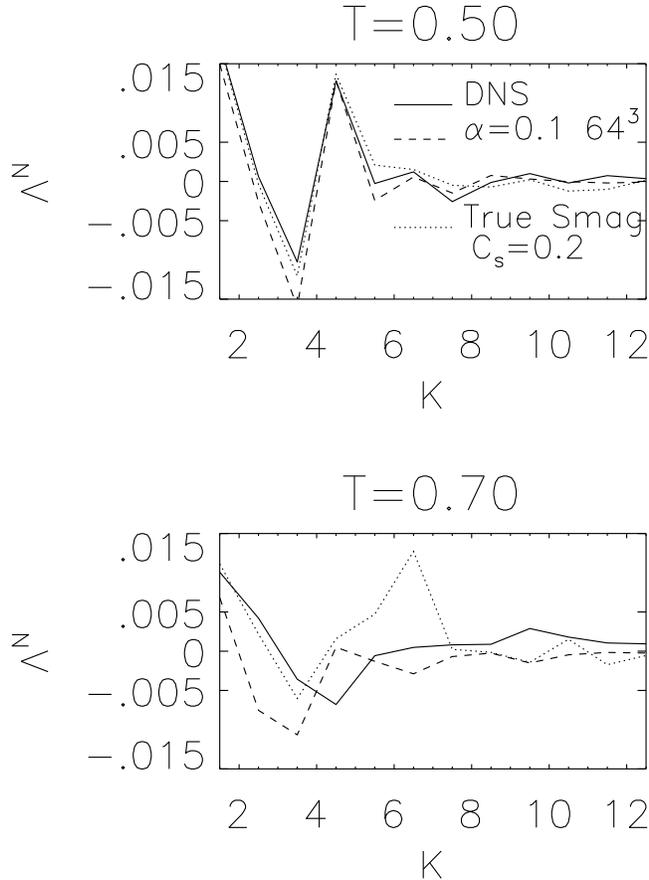,height=12.5cm}
\caption[]{Comparison of the normalized helicity co-spectra 
$\Lambda^N(k)$ for two times.  The normalization is based upon the
energy and enstrophy at each time so
that comparisons in magnitudes between the different calculations
can be made.  These strong oscillations first appear
at $t\approx0.45$ when dissipation is still negligible and the
vortex structures are beginning to change from sheets to tubes
via a roll-up instability induced by interactions between
distinct vortex sheets. The higher wavenumber
oscillations disappear as soon as dissipation becomes significant.
}
\label{helicity-spectra}
\end{figure}

\begin{figure}[htbp]
\psfig{file=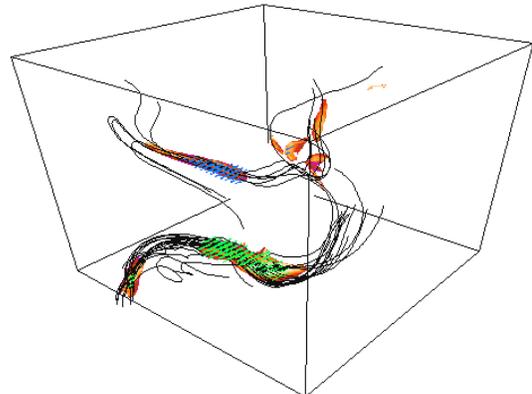,height=6.5cm}
\caption[]{Isosurfaces of vorticity at $t=0.7$ where
$\omega\geq0.47\|\omega\|_\infty$ in red/earth colors
with sample vortex lines through these regions.
Regions of high positive and negative helicity are indicated
by green and blue respectively.
The visualization volume is centered on the new maximum vorticity and
rotated with respect to Fig. \ref{orthogonal-tubes} to emphasize
Beltrami nature of the vortex tubes.  From another angle, it
can been seen that the
transverse alignment found in Fig. \ref{orthogonal-tubes} persists.
}
\label{p7-helicity-renderings}
\end{figure}

While our analysis also includes traditional diagnostics such as energy
decay, our new understanding will arise primarily
through: the time evolution of the maximum vorticity in
Figure \ref{vorticity-time}; physical space visualizations
in Figures \ref{orthogonal-tubes} and \ref{p7-helicity-renderings};
helicity probability distributions in Figure \ref{helicity-pdf};
and the accompanying helicity co-spectrum in Figure \ref{helicity-spectra}.
Central to our new understanding is the following observation: although
this initial condition is not a Beltrami flow,
spatial regions develop early during its evolution where the helical alignment
measured by $\cos\theta=\Lambda/(u\,\omega)$ of either sign is locally
near unity in magnitude.  Evidence is given that the ensuing dynamics is
influenced by helicity $\Lambda$, whose evolution 
depends upon the interplay between nonlinear transport terms,
viscous effects and large-eddy stress parameterizations.

The time scale for this investigation is set by a
surge in the growth of the maximum vorticity at $t=0.5$ in Figure
\ref{vorticity-time}. This time scale also appears in three $64^3$
results: the $256^3$ DNS filtered onto a $64^3$ mesh, in a Smagorinsky
calculation using a traditional eddy viscosity model,
\rem{\cite{Smag[1963]}, }
and in
a calculation using the new LANS-$\alpha$ model, which preserves
nonlinear transport properties
(see \cite{FHT[2001]} and references therein).  The growth
of vorticity comes from the vortex stretching terms, which
at early times involves only large-scale strain and,
thus, should not be affected by the small scales.  Evidence
that vortex stretching at early times
is not strongly affected by the small scales
is that all three calculations
have the same time scale $t=0.5$ for the vorticity surge.
The differences among the calculations should therefore be due
to how vorticity is suppressed by either viscous effects or LES model effects.
The identical growth in $\|\omega\|_\infty$ in each case until $t=0.3$ 
tells us that until that time dissipation and LES parameterization have not 
yet affected the calculations.  

It has previously been shown that the dominant initial structures
are vortex sheets that arise out of non-Beltrami initial conditions,
\cite{Herring-Kerr[1993]}.  From the interaction of those initial vortex
sheets in the weakly  dissipative regime, Fig. \ref{orthogonal-tubes}
shows that a new configuration of transversely aligned vortex structures 
has formed by $t=0.5$.  We will demonstrate that
this is an inherently helical configuration that has
arisen from the non-Beltrami initial conditions 
simultaneously with the vorticity surge event.
The helical nature of the configuration is shown both by the concentrations
of helicity density $\Lambda$ near the vortex structures in Figure
\ref{orthogonal-tubes} and by a skewed, transient probability
density distribution (PDF) of the
cosine of the helicity angle at $t=0.5$ in Figure
\ref{helicity-pdf}.  Accompanying this helicity signature in
physical space, the helicity co-spectrum at low wavenumbers in Figure
\ref{helicity-spectra} develops a strong signature of alternating sign
between wavenumber bands.
This is evidence for the dynamical formation of
large, asymmetrical distributions of helicity associated with
the vorticity surge in Fig. \ref{vorticity-time}.

While the helicity density is not Galilean invariant, 
Galilean transformations cannot remove the strong fluctuations
in helicity in physical space that we observe, 
nor could they remove the strong fluctuations in the helicity 
co-spectrum in Fig. \ref{helicity-spectra}.  The
implication is that if these are configurations that naturally
and frequently arise in a turbulent flow, then helicity could be
playing a central role in their dynamical evolution.
The PDF of the helicity angle in Fig. \ref{helicity-pdf}
quickly changes into a distribution with 
peaks concentrated at $\cos\theta=\pm1$ as
reported in other turbulent flows \cite{Pelzetal85,PolifkeS89}. 
The appearance of the individual peaks is associated with the formation of
nearly Beltrami vortex tubes by $t=0.7$, shown 
Fig. \ref{p7-helicity-renderings}.  


The association of the vorticity surge with helicity raises
some important questions about the absence of anti-parallel vorticity
elements.  In the inviscid limit, 
anti-parallel vorticity elements around the position
of the maximum vorticity $\|\omega\|_\infty$, 
were thought to lead to the strongest increases in
vorticity \cite{Kerr[1993],PumirSiggia[1987]}.  
In a strictly anti-parallel configuration there would be equal concentrations
of both signs of the helicity, which would cancel at each length scale 
and thus preclude any spectral oscillations.

Instead, at no time during this calculation are the the most
intense vorticity elements observed to be anti-parallel.
The calculations
that suggest that the anti-parallel configuration would be dominant
are all based upon inviscid calculations using
vortex filaments and tubes 
(see \cite{PumirSiggia[1987]} and references in \cite{KidaTakaoka[1994]}).
The spectral calculations reported here, which are
viscous and initially develop vortex sheets, show no such trend.

Isolated, helical vortex tubes first form immediately 
after the vorticity surge at $t=0.5$.
The helical character of these vortex tubes is demonstrated by
physical space renderings that show that the strongest helicity 
fluctuations are clearly tied to vortex tubes in
Fig. \ref{p7-helicity-renderings} at $t=0.7$ and by strong peaks
at both 1 and -1 in the helicity distribution in Figure \ref{helicity-pdf}.
This PDF was centered upon the domain shown.  Note that the
two dominant tubes have opposite signs of helicity.  Rotating one's view
of this local region shows that the tubes are nearly orthogonal and
when the entire flow is rendered, this
configuration appears in a localized corner $1/4^3$ of the entire
domain.  This demonstrates that these structures are interacting strongly
within this localized region.  
As time goes on, vortex tubes whose helicity is opposite develop
as orthogonal pairs in many localized regions.
This is not the first time that nearly orthogonal 
configurations of vortex tubes have been created in direct simulations.
They have appeared in renderings going back to about mid-1980s
\cite{Brachetetal[1983],Kerr[1985]\rem{,SheJacksonO[1990],VincentMene[1991]}}.
The new point we are making is that this intrinsically
helical configuration arises in conjunction with the vorticity surge.

The rapid development of structures characteristic of fully-developed
turbulence by $t=0.7$ from the helical state at $t=0.5$
occurs simultaneously with the disappearance of the 
intermediate peak in the helicity co-spectrum in Fig. \ref{helicity-spectra}
seen at $t=0.5$.  The wavenumber ($k=5$) of this peak at $t=0.5$ 
is too low for the disappearance to be a direct effect of viscosity.  
Instead, we believe it is due to a combined effect of transfer to 
larger wavenumbers, followed by dissipation at those larger wavenumbers.  
The evidence for this is the different manner
in which the two large-eddy simulations treat the
nonlinear transfer of helicity.  The LANS-$\alpha$ model
preserves the transport of helicity. Consequently, when
energy is transported to small scales and is dissipated, so should any
helicity that accompanies that energy.  The is consistent with
the observation in Fig. \ref{helicity-spectra}
that the intermediate peak in its helicity co-spectrum
disappears by $t=0.7$.  Graphics also show vortex structures
consistent with the large-scale structures in the DNS.

\begin{figure}[htbp]
\psfig{file=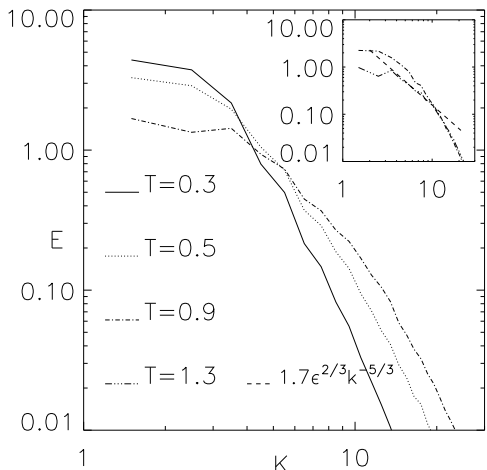,height=6.5cm}
\caption[]{Comparison of the energy spectra versus the
three-dimensional wavenumber by shells for several times.
The main figure shows spectra up to $t=0.9$.
Between $t=0.3$ and $t=0.5$, the high wavenumbers are filled
and there is little decrease in the overall energy.  Between
$t=0.5$ and $t=0.9$, there is a significant decrease in 
the energy in the lowest wavenumbers.
The expected $k^{-5/3}$ law establishes itself by $t=1.3$,
as shown by the inset.  
}
\label{energy-spectra}
\end{figure}

For Smagorinsky, the intermediate peak in the helicity co-spectrum
that had formed by $t=0.5$ persists, 
only moving slightly towards higher wavenumbers by $t=0.7$.
There are no assurances in the Smagorinsky model that both helicity 
and energy will be dissipated. In fact, Smagorinsky can generate
anomalous helicity.  Furthermore, clearly defined vortex tubes
have not formed at this time in the Smagorinsky calculation.
Hence, the proper spectral dynamics of helicity 
and the formation of vortex tubes are part of one nonlinear process
involving diffusion, stretching and proper transport.
This implies that a large-eddy simulation that has correct
transport properties is required to capture this process.

Once the interacting, transversely aligned, helical vortex tubes
have formed, dissipation grows rapidly in the DNS and as has been 
reported\cite{Herring-Kerr[1993]} and shown in Fig. \ref{energy-spectra},
the kinetic energy spectrum $E(k)$ gradually approaches
the classical $k^{-5/3}$ 
characteristic of a turbulent energy cascade.  Once a clear
$-5/3$ spectrum appears near $t=1.3$, the energy decay
becomes self-similar \cite{KidaMurakami[1987]}.

Interactions between regions or modes of oppositely signed
helicity have previously been investigated in the context of 
shell models.  In one of the most popular shell models, the
GOY model, the sign of helicity in each shell alternates.
It has been shown that the GOY interactions are consistent with one channel 
in a helical decomposition of the energy transfer
\cite{Waleffe[1992],BiferaleKerr[1995]}.  
This analysis also shows that the strongest wavenumber interactions occur
between modes of opposite helicity.  In DNS
calculations it has been shown that helicity can arise from
non-helical initial conditions and that energy and helicity
can move together to high wavenumbers and be dissipated
\cite{BiferaleKerr[1995],ChenChenEH[2002]}.  

We conclude that dynamics involving helicity dynamics is essential 
during an early stage in
establishing the turbulent energy cascade from smooth initial conditions.
The new features that we wish to emphasize are the formation
of a helical state in association with the vorticity
surge, and how this state rapidly develops into the structures and
distributions that set the stage for fully developed turbulence.
We believe that understanding this transition could be important
in the parameterization of transient events in turbulence.  Such
transitions must influence the overall statistical properties 
controlled by intermittent, intense events.  
Evidence was presented that the mechanism by which classical
vortex tubes first form requires both transport and dissipation
and that this implies that large-eddy parameterizations should
preserve some of these helicity transport properties.  In ongoing work
we are investigating the capabilities of several modern approaches 
to large-eddy simulation in representing these helicity transport properties.

{\bf  Acknowledgements} We are grateful to many friends for their
enormous help in supplying suggestions, advice and encouragement
during the course of this work. In particular, several helpful
and timely discussions with S. Y. Chen, G. Eyink, U. Frisch and
K. Sreenivasan especially influenced this work.
\vspace{-2mm}

\end{document}